\documentclass[useAMS,usenatbib]{mn2e}
\usepackage{graphicx} 
\title[A Deep 150~MHz GMRT Survey in Eridanus]
{A Deep 150~MHz GMRT Survey in Eridanus}
\author[Samuel J. George, Ian R. Stevens]{Samuel J. George$^{1}$
\thanks{E-mail:samuel@star.sr.bham.ac.uk (SJG)}, Ian R. Stevens$^{1}$
\footnotemark[0]\\
$^{1}$School of Physics and Astronomy, University of Birmingham, 
Edgbaston, Birmingham, UK B15 2TT}
\begin{document}
\date{Accepted **** August **. Received **** December **; in original 
form 2008 July 25th}
\pagerange{\pageref{firstpage}--\pageref{lastpage}} \pubyear{2008}
\maketitle
\label{firstpage}

\begin{abstract}
We present results of a 150~MHz survey of a field centered on
$\epsilon$ Eridani, undertaken with the Giant Metrewave Radio
Telescope (GMRT). The survey covers an area with a diameter of
$\sim
 2^\circ$, has a spatial resolution of $\sim 30''$ and a
noise level of $\sigma=3.1$~mJy at the pointing centre. These
observations provide a deeper and higher resolution view of the
150~MHz radio sky than the 7C survey (although the 7C survey covers a
much larger area).

A total of 113 sources were detected, most are point-like, but 20 are
extended.  We present an analysis of these sources, in conjunction
with the NVSS (at 1.4~GHz) and VLSS (at 74~MHz). This process allowed
us to identify 5 Ultra Steep Spectrum (USS) radio sources that are
candidate high redshift radio galaxies (HzRGs). In addition, we have
derived the $dN/dS$ distribution for these observations and compare
our results with other low frequency radio surveys.

\end{abstract}

\section{Introduction}

Over the past 50 years many large radio surveys have taken place,
allowing us to build up an understanding of the evolution of the radio
source population (see for example, Rowan-Robinson et
al. 1993). Various deep surveys have taken place, but many have been
at frequencies of 1.4~GHz and higher (for obvious angular resolution
reasons) and have reached down to $\mu$Jy levels (e.g. Hopkins et
al. 1998; Huynh et al. 2005). Probably, the most important result from
these surveys is the confirmation of the upturn in the
Euclidean-normalised differential source counts below $\sim 2$~mJy
(Windhorst et. al. 1985) which is believed to be due to a population of
sources that are not AGN (Condon 1989). The ``normal'' population of
radio sources (with flux densities larger than a few mJy) can be
explained by radio loud AGN which are hosted in elliptical
galaxies. Towards $\mu$Jy flux levels the radio sources counts are
increasingly dominated by other populations, such as radio-quiet AGN,
starbursts and late-type galaxies (with their radio emission due to
massive star formation and associated supernovae).

At lower radio frequencies (below 1.4~GHz) the population of radio
sources is less explored, mostly due to the relatively bright flux
limits achieved at these frequencies. The Westerbork Synthesis Radio
Telescope (WSRT) has been used for this purpose reaching a detection
limit of $\sim 2.5$~mJy at 610~MHz, however this is not sufficiently
deep to probe the upturn in source counts (Valentijn 1980). The Giant
Metrewave Radio Telescope (GMRT) located near Pune, India, however,
has the necessary capabilities (low frequency with good sensitivity)
and there have been a few deep surveys using the GMRT at 610~MHz.
Examples include Moss et al. (2007), Garn et al. (2007), Bondi et
al. (2007) and Garn et al. (2008). 

The 150~MHz sky is still less explored. Although the GMRT is capable
of observations at this frequency, there have been relatively few
results presented in the literature. At 150~MHz the GMRT has a spatial
resolution approaching $20''$ and a theoretical {\sl rms} noise of around
$\sim 1$~mJy. This compares favourably with the Cambridge 7C survey
(Hales et al. 2007), which had a resolution of $70\times 70
\csc\delta$ arcsec$^2$, and a typical {\sl rms} noise of around
$\sim 20$~mJy. Of course, the 7C survey covers a much larger area than
is covered in this survey, and the final 7C catalogue contains 43683
sources over an area of 1.7~steradians.  The 7C survey region does not
overlap with the field presented here.

\begin{figure}
\includegraphics[scale=0.3]{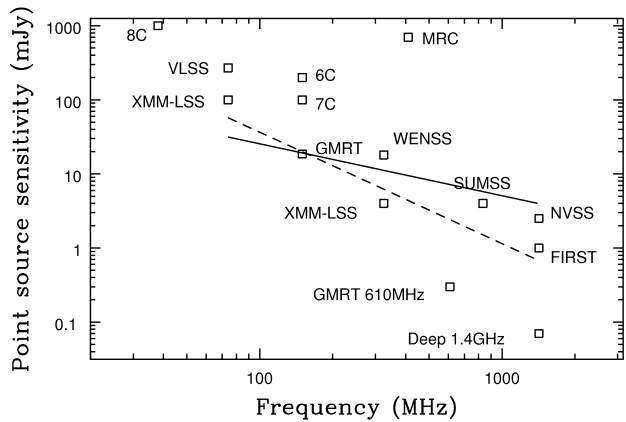}
\caption{A comparison between the point source sensitivity of the 
GMRT 150~MHz results presented here and other low frequency radio 
surveys. To illustrate the respective point source 
sensitivities of the surveys, the two lines are for spectral indices of 
$\alpha=0.8$ (solid line) and $\alpha=1.5$ (dashed line), 
normalised to the sensitivity of the GMRT 150~MHz observations presented 
here. See text for a more detailed description of this diagram.}
\label{surveys}
\end{figure}

A comparison of the point source sensitivities for a range of low
frequency surveys is illustrated in Fig.~\ref{surveys}. The point
labelled ``GMRT'' is the survey presented here. Other surveys included
the various Cambridge surveys (6C, 7C, 8C) at 38~MHz (Rees 1990) and 151~MHz (Hales et al. 2007), the XMM-LSS survey at 74~MHz and 325~MHz (Cohen et al. 2003), 
The VLA Low-frequency Sky Survey (VLSS) at 74~MHz (Cohen et al. 2007), The
Westerbork Northern Sky Survey (WENSS) at 325~MHz (Rengelink et al. 1997), Molonglo Reference
Catalogue (MRC) at 408~MHz (Large et al. 1981), the Sydney University Molonglo Sky Survey
(SUMSS) at 843~MHz (Mauch et al. 2003) and the NVSS and FIRST both at 1.4GHz 
(Condon et al. 1998, White et al. 1997). We also show examples of what can be 
achieved in deeper smaller area surveys; the ``GMRT 610~MHz'' refer to surveys 
such as Garn et al. (2007) and Moss et al. (2007), which can achieve point source 
detections down to around 0.3~mJy, and ``Deep 1.4~GHz'' refers to surveys such as 
Huynh et al. (2005), with point source detections down to $\sim 0.07$~mJy.

We also show lines corresponding to sources with spectral indices of
$\alpha=0.8$ and $\alpha=1.5$, for a flux $S\propto \nu^{-\alpha}$
(which we shall adopt throughout this paper). We can see from
Fig.~\ref{surveys} that the point source sensitivity of this GMRT survey is well
matched to the NVSS. By comparing the sources detected at 150~MHz with
other surveys (such as the NVSS at 1.4~GHz or the VLSS at 74~MHz) it
is possible to identify sources with extreme spectral indices, which
are termed Ultra-Steep Spectrum (USS) sources.  In turn, these USS
sources are candidates for being high-redshift radio galaxies (HzRGs).

Radio sources can be detected out to high redshift
($z>5$). Empirically, it appears that the majority of high redshift
radio sources tend to exhibit steeper radio spectra (see for example,
De Breuck et al. 2000; Miley \& De Breuck 2008). The origin for this
correlation is unclear. Part of it may be due to a Malmquist bias, but
it now seems likely that the higher gas densities in the galaxy
environments at high redshift (constricting the development of radio
bubbles from such AGN) plays a role (see Miley \& De Breuck 2008 for a
discussion of this). Having identified candidate HzRGs via their radio
spectrum it is then necessary to identify the galaxies in the
optical/IR and subsequent determine the redshift. HzRGs seem to have a
relatively small scatter in a plot of $K$-band magnitude versus $z$, and
they seem to be among the most luminous galaxies at high redshift (De
Breuck et al. 2002). Apart from HzRGs there are a number of other sources 
that exhibit steep spectral indices. These include, but are not limited to, 
fossil radio galaxies, cluster halos or pulsars (Parma et al. 2007, Manchester et al. 2005). 

The primary science driver behind these observations was to detect the
magnetospheric radio emission from the exoplanet orbiting $\epsilon$
Eridani. No detection was made, however, tight constraints on the
possible radio emission were determined and these results are
summarised in George \& Stevens (2007). The wide field, good
sensitivity and reasonable spatial resolution of the GMRT 150~MHz
observations does give an excellent view of the radio sky. From these
observations, we show that deep imaging with the GMRT at 150~MHz
enables us to identify candidate high redshift radio sources.  Some
relevant GMRT results at 150~MHz have been published by
Ishwara-Chandra \& Marathe (2007), although it is clear that this is
still an under-exploited capability.

The observations and data analysis are described in Section~2. We
discuss the generation of the catalogue in Section~3 and present an
analysis of the radio source counts at 150~MHz in Section~4. We then
investigate the nature of the sources, including spectral indices and
identify the candidate high redshift objects in Section~5. We
conclude in Section~6.

\section{The GMRT 150~MHz Observations and Data Analysis}

\subsection{Observations}

The radio observations discussed here were conducted with the GMRT,
during cycle 11 on 10 August 2006 and lasted for a duration of
4.13h. The pointing centre was $\epsilon$ Eridani (RA: $03^h 32^m
55^s$ Dec: $-09^\circ 27' 29''$). The central frequency of the
observation was 157.0~MHz and a bandwidth of 8~MHz was used (split
into 128 channels). For flux density calibration 3C48 was observed and
for phase calibration purposes 0114--211 was used. The time on source
in a single scan was 30 minutes before moving to the phase calibrator
for 5 minutes. All reduction was performed with the Astronomical Image
Processing System (AIPS). The visibilities from each baseline and
channel were carefully inspected and edited to remove interference,
cross-talk between antennas, instrumental glitches and the
poor-quality data recorded at the beginning of each pointing was
removed. The flux density scale was set by an AIPS implementation of
the Baars et al. (1977) scale.

\subsection{Imaging Strategy and Calibration}

\begin{figure*}
\includegraphics[scale=0.7]{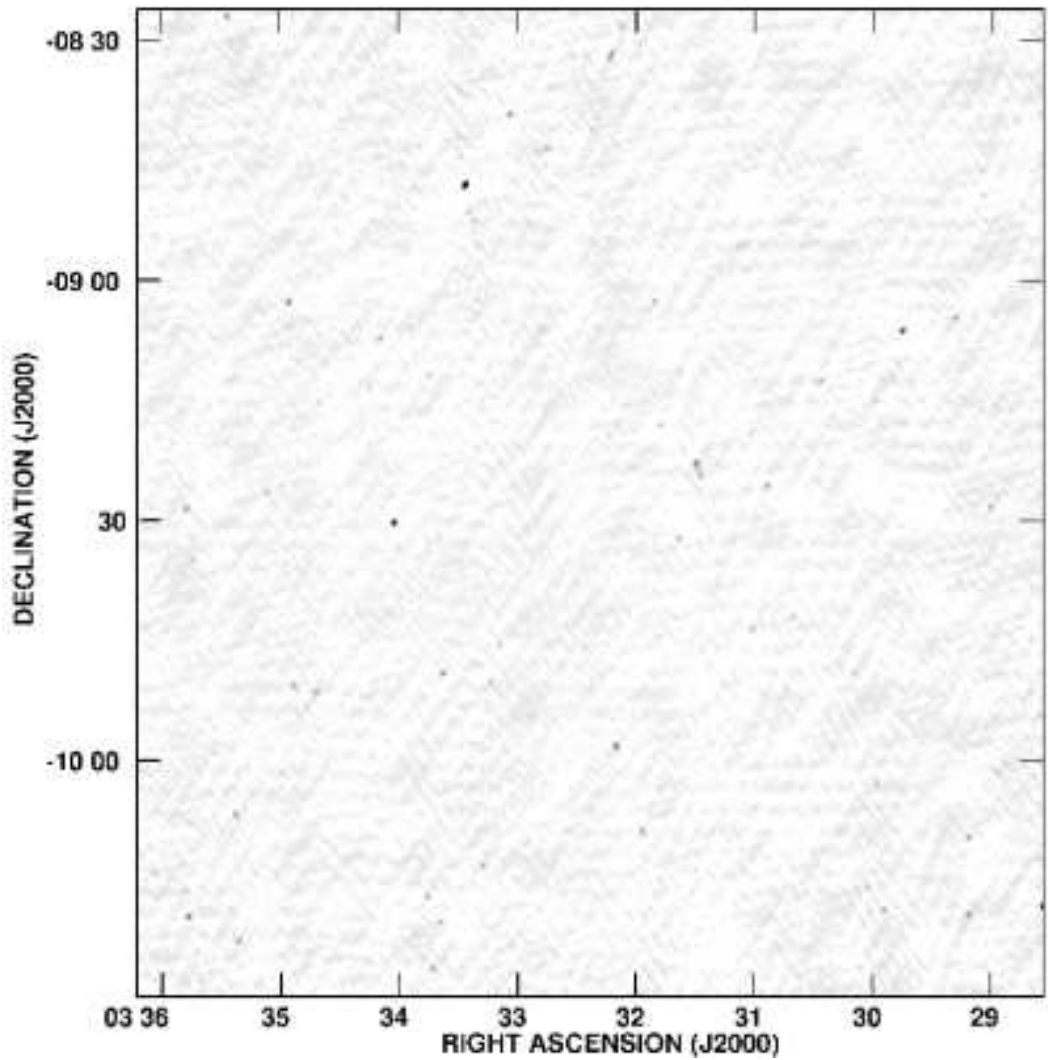}
\caption{The 150~MHz image of the region surrounding $\epsilon$
  Eridani. The field of view considered is around $2^\circ$ in
  diameter, and the noise level is relatively uniform with $\sigma =
  3.1$~mJy. The grey-scale ranges between $0$ and 500~mJy~beam$^{-1}$. 
  For a list of the sources present see Table~1.}
\label{fovimage}
\end{figure*}

Wide-field mapping with a $5^{\circ}$ diameter (close to the full
width of the beam) was performed over 121 facets, with the phase of
the data shifted to the tangent point at the centre of each facet.
This ensures that no point on any image is far from the tangential
pointing position, thus we minimise the smearing of sources far from
the phase tracking centre (due to the breaking of the coplanar array
assumption by using a two-dimensional FFT to approximate the celestial
sphere). Imaging and deconvolution was performed using the AIPS task
{\it IMAGR}. The final 121 image facets used for analysis were total
intensity maps, with a final restoring beam (the resolution) of
$32.1'' \times 23.2''$. The central {\sl rms} ($\sigma$) noise level
achieved was 3.12~mJy, though areas of higher {\sl rms} noise surround 
the sources (with a typical noise of around 5~mJy, rising for the very 
brightest sources) and near the edges of the field. For a given flux level, 
we found a large number of spurious sources in the outer regions and have 
focused in on the central regions of the observation.

This noise level can be compared with the theoretical thermal noise
limit for the GMRT at 150~MHz, given by
\begin{equation}
\sigma = \frac{\sqrt{2} T_{s}}{G \sqrt{n(n-1)N_{IF} \Delta \nu \tau}}\,,
\end{equation}

\noindent where $T_{s}$ is the telescope system temperature (615K),
$G$ is the antenna gain (0.33 K~Jy$^{-1}$), $n$ is the number of
working antennas (typically 27 during our observation), $N_{IF}(= 1)$
is the number of sidebands, $\tau \sim 10^{4}s$ is total time on
source and $\Delta \nu = 5.6$~MHz is the effective bandwidth for each
sideband after channel collapsing\footnote{$http://www.gmrt.ncra.tifr.res.in$}.
The theoretical noise limit of the GMRT is then $\sim 0.4$~mJy/beam.
Our observations are a moderate factor above the theoretical
limit.  The reason for the limitation is probably due to radio frequency 
interference during the observations (particularly low level RFI), though 
limitations from the dynamic range also make a contribution.
The maps of the individual facets were combined into a final
mosaic using the AIPS task {\it FLATN}, with no correction made for
the degradation of the primary beam sensitivity at this stage (any
flux density corrections are considered later).

The final image is shown in
Fig.~\ref{fovimage}. We will discuss the source detection criteria in
the following section.

As we are dealing with low-frequency observations, atmospheric
refraction is likely to be an issue. In order to test the positional
accuracy of our sources we compared the positions of the 20 brightest
objects with their 1.4~GHz counterparts in the NVSS.  On average there
was an offset of $\sim 52''$ between our extracted source positions
and the location of sources in the NVSS. We attribute this offset to
refraction in the Earth's ionosphere. This corresponds to an offset of
$-3.48^{s}$ in RA and $-5.35''$ in Dec and the listed positions in the
source catalogue (Table~\ref{sources}) are corrected for this offset.

\section{Generating the Source Catalogue}

\begin{figure*}
\includegraphics[scale=0.2]{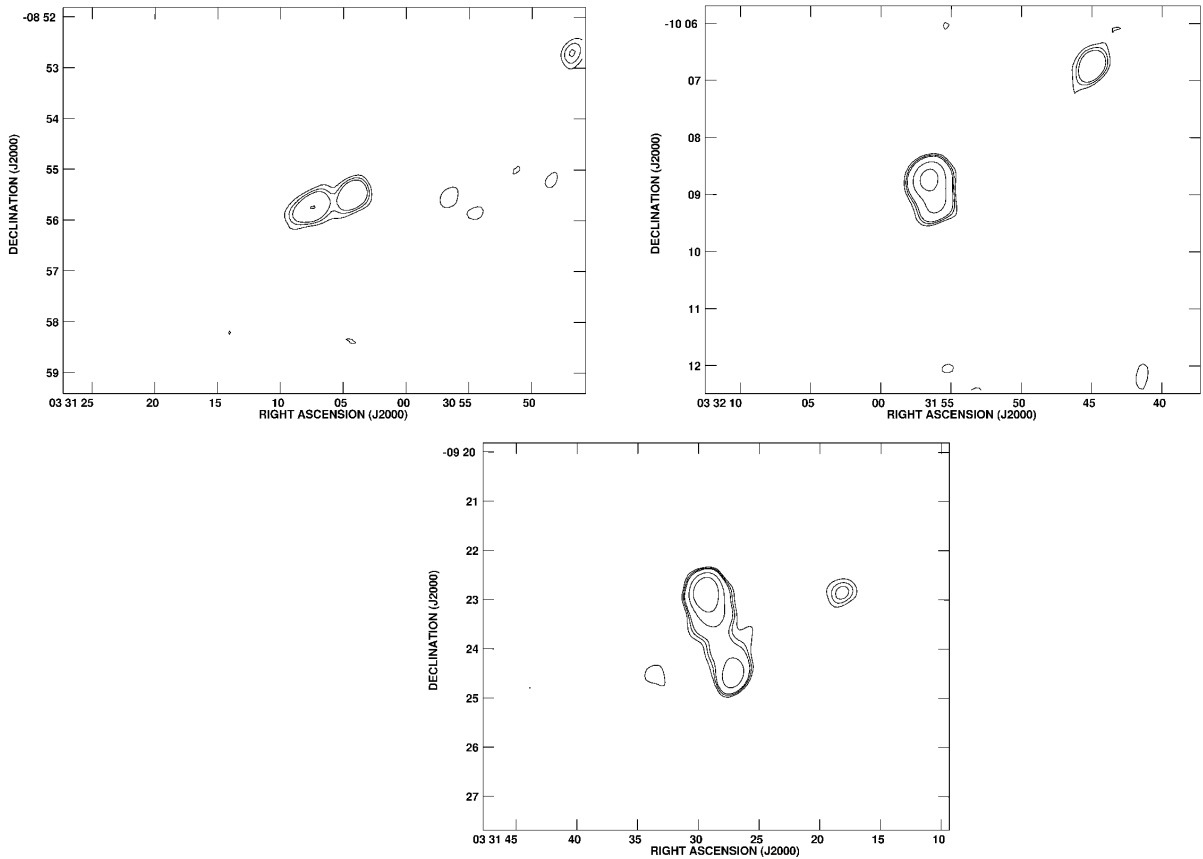}
\caption{A selection of extended objects in the survey, 
GMRTJ033103.8-085539, GMRTJ033152.9-100843, GMRTJ033125.8-092243, 
with two clearly showing double-lobed features. The contours are at 
levels of 10, 15, 20, 50, 100~mJy respectively.}
\label{source_gallery}
\end{figure*}

\subsection{Source Extraction}

The next stage is to construct a source catalogue for these
observations. Because this observation was a single pointing, rather
than a well constructed survey involving a series of pointings to
obtain a fairly uniform sensitivity over a wide area, we have chosen to
consider only the central region of the observation. 

Sources were only included within a 1.07$^{\circ}$ radius from the pointing centre 
(out to the point that the  primary beam correction dropped to 20 per cent of its central value) and allows us to avoid higher noise outer regions of the dataset.
A {\sl rms} noise map of the survey area was
constructed for use within the source detection procedure.
Sources are extracted using the AIPS task {\it SAD}. 
{\it SAD} attempts to find all sources in an image whose peak brightness is 
greater than a given level and then by fitting Gaussian models to the image 
by least squares method. During the detection process the {\sl rms} is 
found by simply looking up the coordinate of the component in the {\sl rms} noise map.
In this region a conservative peak brightness detection limit of 
6$\sigma$ (corresponding to a flux limit of
18.6~mJy) was used to minimise the number of noise spikes spuriously
detected as sources. In the regions around the brightest sources the
detection was performed separately using higher thresholds and visual
inspection.

Within this region a total of 113 sources were found above the
6$\sigma$ threshold. Any extended sources for which the Gaussian model used by {\it SAD}
inadequately described the data (i.e. sources with angular sizes greater than $90''$ or showing some form of substructure upon visual inspection), were inspected using the
AIPS task {\it TVSTAT}. There were 9 sources that this procedure was
used for, at least 2 of these sources appear to be clear double-lobed
sources.  In Fig.~\ref{source_gallery} we present a small selection of
these extended sources.

\subsection{Resolution Effects}

\begin{figure}
\includegraphics[scale=0.4]{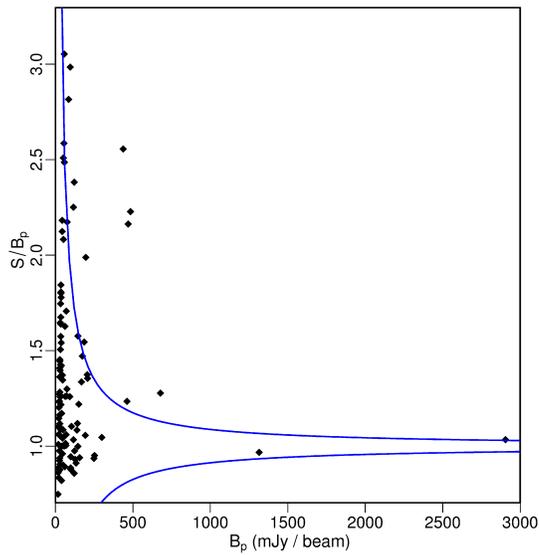}
\caption{The ratio of the flux density to peak brightness $(S/B_{p})$ 
as a function of the peak brightness ($B_{p}$) for the 113 GMRT 
sources. The solid lines indicate the upper and lower envelopes of the flux ratio
distribution used to define an unresolved source.}
\label{totaltopeak}
\end{figure}

When analysing low signal-to-noise ratio detections, noise spikes can
easily cause Gaussian fitting routines (such as {\it SAD}) to poorly
fit the width of a source, giving inaccurate measures of the flux density. 
To investigate this effect, we plot the ratio between
the flux density, $S$, and the peak brightness, $B_{p}$, as a
function of the peak brightness for all the radio sources in the
catalogue. Under the assumption that the measurement of any source
that has a flux density less than the peak brightness is due to the effect of
noise, we use the distribution of the ratio to define a criterion for
a source to be considered resolved. This distribution is then mirrored
above the $S/B_{p} = 1$ value and can be seen in
Fig.~\ref{totaltopeak}. Any source within this envelope is considered
as unresolved and its peak brightness is taken as the best measure
of its true flux density. The lower envelope contains 98$\%$ of the
sources with $S < B_{p}$. The upper envelope is
characterised by the equation

\begin{equation}
\label{res_envelope}
\frac{S}{B_{p}} = 1 + \left( \frac{28}{(B_{p} / \sigma)} \right)\,.
\end{equation}

Using the above envelope a total of 20 sources (18$\%$) are considered to be adequately
resolved and we take their flux density as a measure of their true flux.

\subsection{Flux Density Corrections}

We must first consider the effect of bandwidth smearing on our data. 
This is seen as a radial blurring of the source. The flux density of 
a source is conserved but the peak brightness is reduced and this
effect is the result of many channels of finite bandwidth being used
in the interferometric inversion, that assumes monochromatic
radiation. To minimise this effect we chose to avoid channel
collapsing over a large number of channels. The total bandwidth over
all the channels was 8~MHz and so, to avoid this effect, we collapsed
our 128 channels (each of 62.5~kHz) to 11 channels (each of
687~kHz). This allows us to improve the noise while avoiding the
effect of bandwidth smearing.

Another important effect is time-delay smearing, due to the rotation
of source in the sky during the integration used and this is avoided
by taking short integrations. For our observations we used an
integration time of 8s.

\begin{figure}
\includegraphics[scale=0.4]{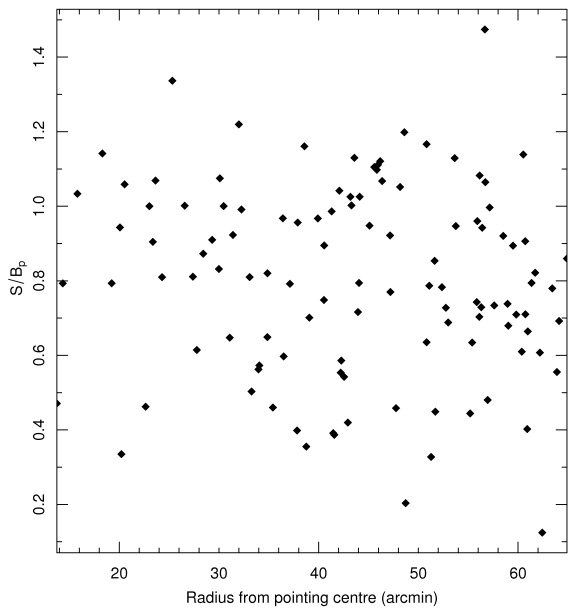}
\caption{The variation of the ratio of the flux density to peak brightness $(B_{p}/S)$ 
across the image for the 113 GMRT sources. If significant bandwidth 
or time smearing was present then a reduction of $B_{p}/S$ with 
distance from the pointing centre of the observations would be seen, this is not the case. }
\label{variationofS_frac}
\end{figure}

Fig.~\ref{variationofS_frac} shows the ratio of measured peak brightness to
measured flux density of all the observed sources. If either of
the above mentioned smearing effects were important it would be
manifested in this plot by a clear relationship with radial distance
from the pointing centre (i.e. the phase tracking centre).  We can
confirm that neither case presents a significant effect on the quality
of our data by running a Pearson's correlation test on the sources.
The correlation test gives a value of $-0.10$ suggesting that there is
no significant correlation. We do not consider these effects further.

Sources that are far from the pointing centre of the observations 
could have their flux smeared out due to the assumption
of a coplanar array. The effect was minimised by our imaging strategy
of splitting the data into many smaller facets.

Importantly, any interferometric observation is modulated by the
primary beam pattern of the array elements. The flux densities of all
objects were corrected for the GMRT 150~MHz primary beam response
using an eighth-order polynomial provided by Kantharia \& Pramesh Rao
(2001).

\subsection{The GMRT 150~MHz Eridanus Field Catalogue}
\label{cat}

These procedures have produced a final catalogue of 113 sources.  The
full source catalogue is given in Table~\ref{sources} and is presented
in order of increasing RA. In this table the columns are: (1) GMRT
Source Name; (2) RA (J2000); (3) Dec (J2000); (4) peak brightness,
$B_{p}$ in mJy~beam$^{-1}$; (5) flux density, $S$ in mJy, 
for resolved sources; (6) 1$\sigma$ error on $B_{p}$ for point sources 
or $S$ for resolved sources; (7) angular size (diameter) in 
arcsec (determined from the deconvolution of the clean beam by {\it SAD}, 
apart from 9 targets which were extracted with {\it TVSTAT} tool and in these 
cases we give the size as the line of greatest extent)
for all sources including those not resolved. The sources with parameters
determined by {\it TVSTAT} rather than {\it SAD} have an asterisk
after the value; (8) the peak NVSS (1.4~GHz) flux (if the GMRT source
has a counterpart) (9) the peak VLSS (74~MHz) flux.

\begin{table*}
\setcounter{table}{0}
\centering
\caption{The complete GMRT 150~MHz source catalogue, listed in order
  of RA. 
GMRT Source Name, RA, Dec (J2000.0), peak brightness, $B_{p}$, in mJy~beam$^{-1}$, 
flux density (only given for resolved sources), $S$ in mJy, 
the 1$\sigma$, error on $B_{p}$ for point sources or $S$ 
for those resolved, angular size in arcsec (see Section~\ref{cat}) 
and for the sources with NVSS(1.4~GHz)/VLSS(74~MHz) counterparts 
the peak flux from these surveys. The * denotes that the source flux
was determined by manual inspection using the AIPS task {\it TVSTAT}. 
See text for further details of these parameters.}
\label{sources}\smallskip
\begin{tabular}{cccccccccc}
\hline
Source Name & RA & DEC & $B_{p}$ & $S$ & $\sigma$ & $\theta$ & NVSS & VLSS \\
    & & & & & & & Peak Flux & Peak Flux\\
    & (J2000) & (J2000) & (mJy beam$^{-1}$) &  (mJy) &(mJy) &  (arcsec) &  (mJy) & (mJy) \\
\hline
GMRTJ032832.5-092649 & 03 28 32.55& --09 26 49.4 & 24.2  & & 5.6 & 54.4&&\\ 
GMRTJ032851.6-094857 & 03 28 51.63& --09 48 57.2 & 20.4  & & 5.4 & 63.4&&\\ 
GMRTJ032856.6-092809 & 03 28 56.68& --09 28 09.4 & 165.8 & & 5.7 & 37.7 &54.2&\\ 
GMRTJ032915.1-090435 & 03 29 15.10& --09 04 35.5 & 120.3 & 185.8 & 13.0& 51.0 &35.5&\\ 
GMRTJ032941.6-090614 & 03 29 41.60& --09 06 14.3 & 530.5 & 677.8 & 11.6&43.7&147.5&1130\\ 
GMRTJ032953.9-100254 & 03 29 53.93& --10 02 54.8 & 73.7  & & 11.5 & 43.3&226.7&\\ 
GMRTJ032955.4-091457 & 03 29 55.45& --09 14 57.5 & 68.0  & & 5.7 & 40.2&9.2&\\ 
GMRTJ033005.4-094855 & 03 30 05.41& --09 48 55.4 & 50.1  & & 5.7 & 45.1&13.0&\\ 
GMRTJ033016.9-091649 & 03 30 16.98& --09 16 49.0 & 26.9  & & 5.7 & 36.3&2.5&\\ 
GMRTJ033022.8-091233 & 03 30 22.84& --09 12 33.8 & 126.5 & & 5.7 & 38.9&34.3&\\ 
GMRTJ033037.0-094153 & 03 30 37.07& --09 41 53.7 & 64.1  & & 10.0 & 39.9&6.1&\\ 
GMRTJ033043.1-085236 & 03 30 43.16& --08 52 36.7 & 20.1  & & 5.6 & 47.1&4.2&\\ 
GMRTJ033049.8-092538 & 03 30 49.88& --09 25 38.7 & 123.5 & & 11.3 & 42.6&42.1&\\ 
GMRTJ033051.4-095140 & 03 30 51.44& --09 51 40.0 & 23.5  & & 5.7 & 43.7&3.2&\\ 
GMRTJ033055.5-100556 & 03 30 55.51& --10 05 56.6 & 27.1  & & 5.7 & 39.7&14.7&\\ 
GMRTJ033057.2-091847 & 03 30 57.22& --09 18 47.2 & 47.7  & & 9.7 & 41.0&10.0&\\ 
GMRTJ033057.5-094330 & 03 30 57.59& --09 43 30.5 & 94.0  & & 5.7 & 41.3&18.1&\\ 
GMRTJ033102.9-085535 & 03 31 02.97& --08 55 35.2 & 40.9  & 69.8 * & 5.2&123.9&9.9$^{\dagger}$& \\ 
GMRTJ033103.8-085539 & 03 31 03.85& --08 55 39.3 & 51.0  & 121.5 * & 5.7&107.7&9.9$^{\dagger}$&\\
GMRTJ033114.7-092245 & 03 31 14.77& --09 22 45.9 & 23.2  & & 5.6 & 46.3&2.9&\\ 
GMRTJ033125.8-092243 & 03 31 25.87& --09 22 43.7 & 217.5 & 484.7 *&5.7&157.6&89.8&\\
GMRTJ033129.3-084116 & 03 31 29.37& --08 41 16.6 & 22.1  & & 5.7&46.9&5.9&\\ 
GMRTJ033134.7-093213 & 03 31 34.71& --09 32 13.3 & 111.2 & & 5.7&36.9&41.5&\\ 
GMRTJ033141.5-100639 & 03 31 41.55& --10 06 39.3 & 45.3  & & 5.7&42.2&11.1&\\ 
GMRTJ033143.7-091802 & 03 31 43.74& --09 18 02.7 & 67.9  & & 9.2&36.4&22.6&\\ 
GMRTJ033146.7-090235 & 03 31 46.75& --09 02 35.5 & 141.2 & & 9.0&37.2&12.8&\\ 
GMRTJ033146.8-102147 & 03 31 46.83& --10 21 47.2 & 26.4  & & 5.7&38.4&5.5&\\ 
GMRTJ033149.4-083344 & 03 31 49.35& --08 33 44.3 & 28.5  & & 5.7&39.1&8.9&\\ 
GMRTJ033151.5-085303 & 03 31 51.50& --08 53 03.4 & 20.5  & & 5.7&49.8&2.1&\\ 
GMRTJ033152.9-100843 & 03 31 52.97& --10 08 43.7 & 127.2 & 142.2 *&5.5&95.8&37.0&950\\
GMRTJ033154.3-082506 & 03 31 54.32& --08 25 06.7 & 20.0  & & 5.4&60.7&&\\ 
GMRTJ033203.0-082817 & 03 32 03.08& --08 28 17.9 & 112.9 & & 5.7&42.6&33.6&\\ 
GMRTJ033205.1-082743 & 03 32 05.19& --08 27 43.6 & 22.1  & & 5.6&59.1&&\\ 
GMRTJ033206.4-095810 & 03 32 06.41& --09 58 10.3 & 374.2 & 462.0&11.4&42.9&75.9&800\\ 
GMRTJ033209.1-083151 & 03 32 09.18& --08 31 51.8 & 171.1 & 437.4&5.4&84.2&91.5&\\ 
GMRTJ033209.4-095929 & 03 32 09.43& --09 59 29.7 & 19.1  & & 5.7&44.0&3.1&\\ 
GMRTJ033210.3-091926 & 03 32 10.30& --09 19 26.1 & 20.5  & & 5.7&40.0&5.6&\\ 
GMRTJ033213.5-094923 & 03 32 13.59& --09 49 23.1 & 21.5  & & 5.6&52.4&&\\ 
GMRTJ033215.3-085907 & 03 32 15.34& --08 59 07.2 & 19.0  & & 5.7&47.9&108.7&\\ 
GMRTJ033216.0-083940 & 03 32 16.06& --08 39 40.6 & 21.4  & 55.2 &5.4&65.7&3.3&\\ 
GMRTJ033218.6-084106 & 03 32 18.67& --08 41 06.3 & 57.8  & & 5.7&43.6&18.9&\\ 
GMRTJ033221.7-094548 & 03 32 21.70& --09 45 48.4 & 32.7  & & 5.7&37.0&6.9&\\ 
GMRTJ033228.4-083154 & 03 32 28.44& --08 31 54.2 & 48.2  & & 5.7&36.4&8.9&\\ 
GMRTJ033234.6-091403 & 03 32 34.62& --09 14 03.3 & 29.4  & & 5.7&40.4&5.8&\\ 
GMRTJ033237.6-082551 & 03 32 37.68& --08 25 51.6 & 30.2  & 85.1&5.5&54.0&7.2&\\ 
GMRTJ033238.9-083552 & 03 32 38.91& --08 35 52.6 & 18.7  & 57.2 &5.3&75.8&&\\ 
GMRTJ033241.2-084330 & 03 32 41.23& --08 43 30.4 & 92.3  & & 5.6&49.4&34.9&\\ 
GMRTJ033247.3-084553 & 03 32 47.33& --08 45 53.3 & 21.8  & & 5.6&52.6&&\\ 
GMRTJ033255.3-084451 & 03 32 55.37& --08 44 51.4 & 19.7  & & 5.7&47.8&&\\ 
GMRTJ033300.0-083915 & 03 33 00.00& --08 39 15.4 & 265.0 & & 9.6&38.6&31.5&\\ 
GMRTJ033305.3-094533 & 03 33 05.34& --09 45 33.6 & 34.8  & & 5.6&47.7&5.5&\\ 
GMRTJ033305.9-102345 & 03 33 05.99& --10 23 45.5 & 20.6  & & 5.6&55.0&&\\ 
GMRTJ033310.0-095010 & 03 33 10.03& --09 50 10.1 & 57.0  & & 5.7&47.7&23.7&\\ 
GMRTJ033313.5-101302 & 03 33 13.52& --10 13 02.1 & 149.5 & 205.5 *&5.7&110.8&24.8&\\
GMRTJ033313.6-083435 & 03 33 13.61& --08 34 35.9 & 19.7  & & 5.6&50.1&&\\ 
GMRTJ033315.7-085548 & 03 33 15.79& --08 55 48.2 & 51.6  & 116.2 *&5.6&49.9&&\\ \hline
\end{tabular}
\end{table*}

\begin{table*}
\setcounter{table}{0}
\centering
\caption{- continued \label{sources1} }\smallskip
\begin{tabular}{cccccccccc}
\hline
Source Name & RA & DEC & $B_{p}$ & $S$ & $\sigma$& $\theta$ & NVSS & VLSS \\ 
&&&&&&& Peak Flux & Peak Flux\\
 & (J2000)&(J2000)& (mJy beam$^{-1}$) &(mJy) & (mJy) & (arcsec)& (mJy)& (mJy)\\ \hline
GMRTJ033316.6-085256 & 03 33 16.65& --08 52 56.5 & 25.3  & & 5.7&46.5&&\\
GMRTJ033317.9-101015 & 03 33 17.96 & --10 10 15.4 & 26.9 &&5.6 &46.3 &7.2&\\ 
GMRTJ033318.2-085354 & 03 33 18.25 & --08 53 54.6 & 20.8 &&5.7&33.0&&\\ 
GMRTJ033318.7-085227 & 03 33 18.72 & --08 52 27.2 & 39.7 &&5.7&42.8&&\\ 
GMRTJ033320.6-085940 & 03 33 20.67 & --08 59 40.0 & 19.1 &&5.5&53.0&&\\ 
GMRTJ033322.7-084803 & 03 33 22.73 & --08 48 03.1 & 2810.0&2905.0&10.00&40.3 &596.5&3990\\ 
GMRTJ033323.0-084220 & 03 33 23.03 & --08 42 20.1 & 34.6&&5.7&43.9 &&\\ 
GMRTJ033324.8-082629 & 03 33 24.83 & --08 26 29.1 & 24.8&&5.4&61.8 &&\\ 
GMRTJ033325.4-083618 & 03 33 25.49 & --08 36 18.7 & 35.0&&5.4&63.9 &&\\ 
GMRTJ033326.6-084205 & 03 33 26.61 & --08 42 05.9 & 56.4&&10.0&39.1&&\\ 
GMRTJ033329.3-083121 & 03 33 29.32 & --08 31 21.1 & 21.0&&5.7&42.5 &&\\ 
GMRTJ033332.1-083707 & 03 33 32.17 & --08 37 07.5 & 20.6&&5.7&39.8 &&\\ 
GMRTJ033334.0-094904 & 03 33 34.09 & --09 49 04.1 & 265.7&&5.7&38.3&71.2&\\ 
GMRTJ033335.1-102010 & 03 33 35.10 & --10 20 10.3 & 111.5&&5.7&36.6&21.2&\\ 
GMRTJ033339.4-102555 & 03 33 39.44 & --10 25 55.1 & 128.2&&5.7&40.6&45.3&\\ 
GMRTJ033341.0-091146 & 03 33 41.07 & --09 11 46.0 & 55.6&&5.7&38.6&30.2&\\ 
GMRTJ033341.9-101657 & 03 33 41.90 & --10 16 57.0 & 145.0&&10.0&39.2&34.4&\\ 
GMRTJ033343.8-102442 & 03 33 43.87 & --10 24 42.9 & 32.6&&5.6&46.0 & 7.9&\\ 
GMRTJ033344.9-095157 & 03 33 44.96 & --09 51 57.3 & 18.8&&5.7&47.5&&\\ 
GMRTJ033347.3-082516 & 03 33 47.34 & --08 25 16.7 & 23.1&&5.4&80.1&&\\ 
GMRTJ033348.2-102144 & 03 33 48.25 & --10 21 44.7 & 37.6&&5.5&50.9 &14.7&\\ 
GMRTJ033358.7-093013 & 03 33 58.79 & --09 30 13.1 & 1359.0&&10.0&38.9&205.5&1970\\ 
GMRTJ033405.6-090713 & 03 34 05.65 & --09 07 13.2 & 146.5&&9.4&38.0&26.7&\\ 
GMRTJ033406.2-084425 & 03 34 06.20 & --08 44 25.8 & 30.7&&5.7&41.7 &7.7&\\ 
GMRTJ033411.4-091323 & 03 34 11.48 & --09 13 23.5 & 21.1&&5.7&42.9&2.7&\\ 
GMRTJ033412.9-085353 & 03 34 12.97 & --08 53 53.1 & 20.8&&5.5&56.1&5.3&\\ 
GMRTJ033422.4-095133 & 03 34 22.48 & --09 51 33.4 & 19.1&&5.6&47.2&7.6&\\ 
GMRTJ033423.1-090733 & 03 34 23.16 & --09 07 33.7 & 19.8&&5.5&66.4 &&\\ 
GMRTJ033436.9-091515 & 03 34 36.98 & --09 15 15.7 & 43.4&&5.7&39.9 &7.0&\\ 
GMRTJ033438.3-095125 & 03 34 38.26 & --09 51 25.3 & 125.8&&5.6&45.1&25.8&\\ 
GMRTJ033445.7-100132 & 03 34 45.70 & --10 01 32.7 & 19.6&&6.0&51.9 &2.2&\\ 
GMRTJ033449.7-095033 & 03 34 49.71 & --09 50 33.7 & 118.1&&6.0&47.1 &46.7&\\ 
GMRTJ033451.9-090239 & 03 34 51.99 & --09 02 39.7 & 286.3&&10.3&40.8 &29.4&\\ 
GMRTJ033455.7-100127 & 03 34 55.77 & --10 01 27.5 & 29.1&&5.7&44.7 &5.1&\\ 
GMRTJ033501.8-090821 & 03 35 01.84 & --09 08 21.7 & 21.0&&5.5&49.9 &4.8&\\ 
GMRTJ033503.0-092629 & 03 35 03.00 & --09 26 29.9 & 137.1&127.5 *&9.6&105.6&25.5&\\
GMRTJ033511.5-083502 & 03 35 11.55 & --08 35 02.6 & 19.2 &154.6&5.2&125.6&&\\ 
GMRTJ033514.3-095502 & 03 35 14.35 & --09 55 02.8 & 22.2 & &5.7 & 34.3 &4.1&\\ 
GMRTJ033518.1-095050 & 03 35 18.18 & --09 50 50.7 & 20.0 & &5.5 &54.8&&\\ 
GMRTJ033518.9-100638 & 03 35 18.91 & --10 06 38.3 & 153.1& &5.6&45.0&58.1&\\ 
GMRTJ033520.1-091211 & 03 35 20.19 & --09 12 11.9 & 32.0&95.6&5.3&95.5&16.3&\\ 
GMRTJ033521.1-101832 & 03 35 21.15 & --10 18 32.6 & 19.4&&5.5 &54.5&&\\ 
GMRTJ033529.0-090706 & 03 35 29.01 & --09 07 06.6 & 18.7&&5.7 &43.3&&\\ 
GMRTJ033540.4-093452 & 03 35 40.47 & --09 34 52.9 & 65.2&&5.7 &41.1&14.2&\\ 
GMRTJ033542.9-100740 & 03 35 42.96 & --10 07 40.4 & 20.8&&5.5 &50.5&4.9&\\ 
GMRTJ033543.8-092829 & 03 35 43.81 & --09 28 29.3 & 182.5&&5.7 &40.0&33.5&\\ 
GMRTJ033559.1-100610 & 03 35 59.12 & --10 06 10.8 & 20.4&&5.4 & 71.3&&\\ 
GMRTJ033600.2-085417 & 03 36 00.24 & --08 54 17.9 & 32.8&&5.7 & 37.9&8.7&\\ 
GMRTJ033620.6-091844 & 03 36 20.60 & --09 18 44.6 & 18.9&&5.5 &51.5&&\\ 
GMRTJ033626.4-091345 & 03 36 26.49 & --09 13 45.6 & 217.0&469.4 *&5.7&38.9&29.7&\\
GMRTJ033633.9-085921 & 03 36 33.98 & --08 59 21.1 & 19.3&&5.7 &41.4&3.7&\\ 
GMRTJ033637.7-091907 & 03 36 37.78 & --09 19 07.5 & 98.7&196.3 *&9.7 &136.6&16.1&\\
GMRTJ033639.0-092343 & 03 36 39.04 & --09 23 43.3 & 39.2&&5.7 &39.2 &6.2&\\ 
GMRTJ033642.4-092228 & 03 36 42.46 & --09 22 28.3 & 29.3&&5.7&43.4&8.3&\\ 
GMRTJ033644.7-095021 & 03 36 44.71 & --09 50 21.1 & 23.4&&5.7 &44.0&&\\ 
GMRTJ033648.7-094602 & 03 36 48.70 & --09 46 02.0 & 21.0&21.0&5.2 &105.7&&\\ 
GMRTJ033701.7-092958 & 03 37 01.70 & --09 29 58.7 & 104.4&&9.6&38.1&15.7&\\ \hline
\end{tabular}
\begin{list}{}{}
  \item[$^{\dagger}$] These sources appear to be distinct objects in the GMRT observations but appear as one unresolved source in the NVSS.
\end{list}
\end{table*}

\section{Source Counts}

The 1.4~GHz source counts have been heavily studied (for example, Windhorst et al. 1985; 
Hopkins et al. 1998) and show a flattening of the
Euclidean-normalised counts below $\sim$ 1~mJy. At 610~MHz this effect
is also seen (for example, Moss et al. 2007; Garn et al. 2007, 2008).

For a source with a spectral index of $\alpha=0.8$ (with $S \propto
\nu^{-\alpha}$), a 1~mJy source at 1.4~GHz, corresponds to a 6~mJy
source at 150~MHz. Consequently, these 150~MHz observations do not
reach into this `sub-mJy' regime, so we cannot investigate this
population at 150~MHz.

\subsection{Completeness}

Two issues need to be considered when interpreting the measured source
count distribution.

The first is Eddington bias, where random measurement errors result in
more objects being boosted into a given bin of flux density from the
one below than are removed from that bin by having a measurement flux
density lower than the value (provided that the source counts have the
correct sign of slope). 

We derive radio source counts using only sources with flux densities greater 
than $18.6$~mJy. This allows us to minimise the effect of the Eddington bias 
in the lowest flux density bin. Also, at 610~MHz, Moss et al. (2007) found
that this only effected their faintest flux density bin, increasing it
by approximately 20$\%$. Since we have comparatively low source counts
in our smallest bin we choose to make no correction for this effect.

The second effect is the incompleteness to extended sources, a
so-called resolution bias. This correction accounts for the fact that
extended objects with peak brightnesses below the survey limit, but 
flux densities above this limit would not be detected
by our source detection procedure. This can be estimated with knowledge of the true
source angular size distribution as a function of flux density. At
1.4~GHz Windhorst et al. (1990) assume an exponential form for the
integral angular size distribution. This is not necessarily the case
at low frequencies and there is also discussion that this is not the
correct form to take even at 1.4~GHz (Bondi et al. 2003). Little is
known about the angular size distribution of mJy sources at low
frequencies, due to the poor angular resolution that can be achieved
at low frequencies. GMRT 150~MHz observations currently represent the highest
spatial resolution that can be achieved at such low frequencies.  Moss
et al. (2007) use the Bondi et al. (2003) formulation to estimate that
they will miss approximately $3\%$ of the sources due to the
resolution bias at 610~MHz. Here, we expect to miss a negligible
number of extended objects.

\begin{figure}
\includegraphics[scale=0.4]{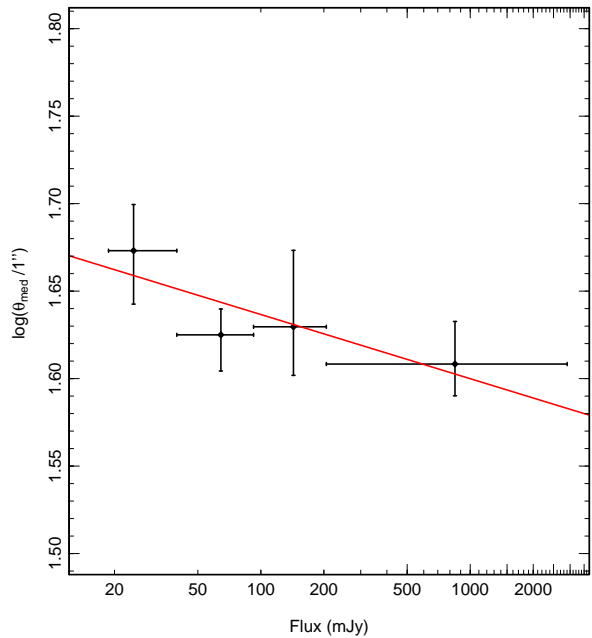}
\caption{The median size versus flux for the GMRT sources. The 113
  sources are binned into 4 bins. The median size is nearly
  independent of flux, the best fit (which is shown) has a gradient of $-0.037$. 
  The $y$-error bars are given at the $35\%$ and $65\%$ angular size. 
  The $x$-error bars are the radial widths of the flux bins. }
\label{mediansize}
\end{figure}

\begin{figure*}
\includegraphics[scale=0.6]{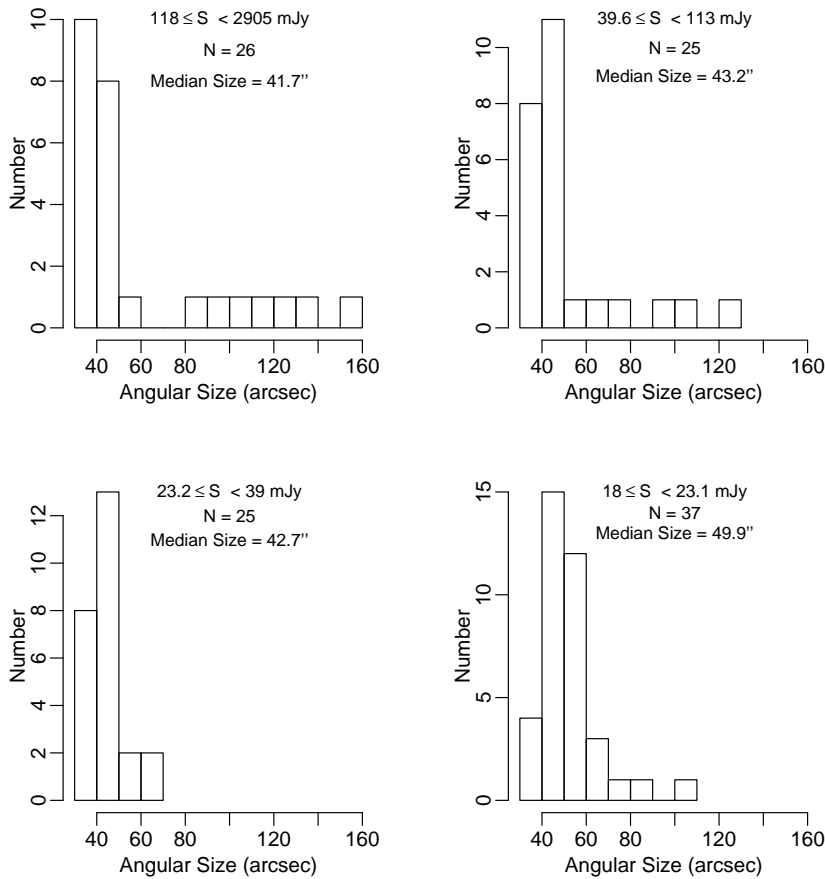}
\caption{The angular size distribution of the GMRT sources. The 
113 sources have been divided into 4 flux bins with roughly equal
number of sources, with the flux range listed in each panel. The low resolution
of the survey means that the angular size distribution 
is peaked around the median angular size.}
\label{histangsize}
\end{figure*}

To illustrate this, in Fig.~\ref{mediansize} we divide the sources up
into 4 flux bins and plot the median source size ($\theta_{med}$) at a
given flux density. The error bars on the median represent the 35\%
and 65\% levels within each flux bin (see Oort 1988). The correlation
is relatively flat, with a gradient of $-0.037$. Further to this, 
Fig.~\ref{histangsize} shows that the shape of the
angular size distribution (in the same flux bins) is always strongly
peaked around the median angular size, regardless of the flux range.
Thus, we choose to make no correction for the resolution bias. It is also important to
stress that, due to our poor resolution, it is easy to overestimate
the median angular size.

\subsection{Differential Source Counts}

We have constructed sources counts by binning our sources by their
integrated flux density, with bins of 0.3 dex in width. The lower limit
was selected as the 6$\sigma$ limit. 

The area over which a source of a given flux density can be detected,
also known as the visibility area, depends on the source peak flux
density and the homogeneity of the noise distribution. The {\sl rms} noise
in our image is 3.1~mJy at the centre and is a function of the
primary beam response with it increasing towards the edges of the
field, of course local variation caused by bright sources also
increase the noise. When calculating the source counts we correct for
this factor by determining what fraction over the map the source can be 
detected.

The source counts were corrected for the fraction of the image over
which they could be detected, taking into account the increase in
noise near bright source, and other biases (as described above). The
differential source count $dN/dS$ was calculated by dividing $N$, the
number in each bin, by $A\Delta S$, where $A$ is the total area of the
image in steradians, and $\Delta S$ is the width of the flux bin in
Jy, i.e.

\begin{equation}
\frac{dN}{dS} = \frac{N}{A\Delta S}\,.
\end{equation}

\begin{figure}
\includegraphics[scale=0.4]{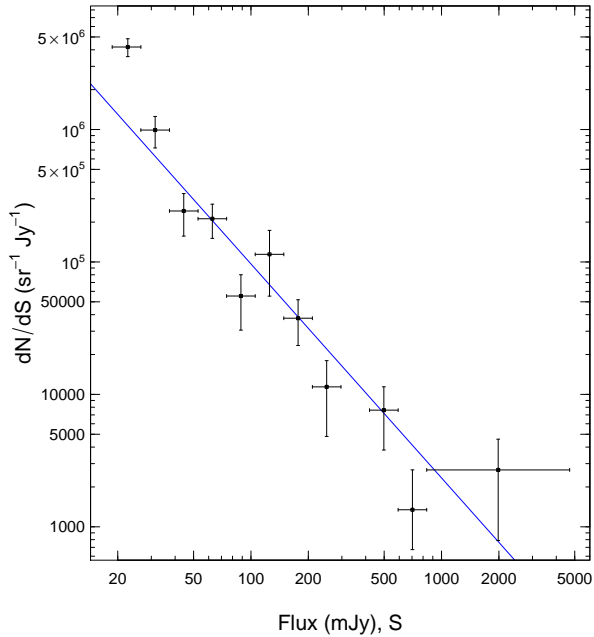}
\caption{The differential sources counts $\left({dN}/{dS}\right)$ 
at 150~MHz from our survey, this shows good agreement with 
the differential sources counts seen at higher 
frequencies (e.g. Garn et al 2007). The $x$-error bars are determined by the width of the flux bins. The fit shown is a power-law with a slope of $-1.61$.}
\label{dnds1}
\end{figure}

\begin{figure}
\includegraphics[scale=0.4]{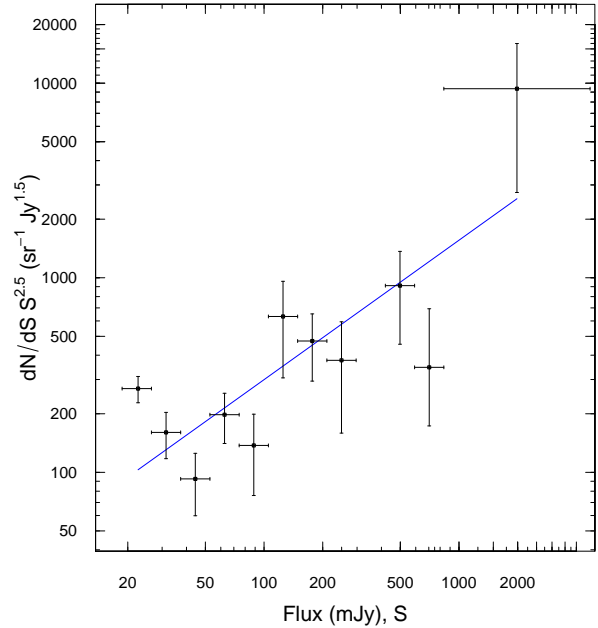}
\caption{Differential sources counts at 150~MHz, normalised by the 
value expected in a static Euclidean universe. A steady decrease 
in the source counts is seen as we approach lower flux. The $x$-error bars are determined by the width of the flux bins. The fit shown is a power-law with a slope of $0.72$.}
\label{dnds}
\end{figure}

Table~\ref{dnds_numbers} gives the flux bin, the mean flux density of
the source in each bin, the number of sources, the corrected number
of, $dN/dS$ and $dN/dS$ normalised by $\left<S\right>^{2.5}$  (the midpoint in the flux bin), which is the
value that is expected from a static Euclidean Universe. A plot of the differential 
sources counts can be seen in Fig.~\ref{dnds1} and shows, as expected, a decrease of 
the differential source counts with increasing flux. For the case of a Euclidean 
Universe a plot of the normalised differential source counts can be found in 
Fig.~\ref{dnds} (and see Table~\ref{dnds_numbers}). This shows a steady increase in 
the source counts as flux increases. The general trend in both of these plots is 
similar to that seen at higher frequencies (e.g. at 610~MHz Garn et al. 2007).
             
\begin{table*}
\centering
\caption{Tabulated 150~MHz differential source counts from this 
survey. The columns show bin flux limits, the bin centre, number 
of sources, the corrected number of sources and Euclidean-normalised 
$dN/dS$. \label{dnds_numbers} }\smallskip
\begin{tabular}{cccccccc}
\hline
Flux Bin & $\left<S\right>$	& $N$ 	& $N_{c}$ 	& $dN_{c}/dS$  & $dN_{c}/dS$ $\left<S\right>^{2.5}$ \\
 (mJy)   &(mJy)		&   	&          	& (sr$^{-1}$Jy$^{-1}$)& (sr$^{-1}$Jy$^{1.5}$) \\
\hline
18.7  -- 26.4   & 22.6   & 42  & 98.8 & $4.20 \pm 0.65 \times 10^6$ & $2.69 \pm 0.42 \times 10^2$\\
26.4  -- 37.4   & 31.4   & 14  & 32.9 & $9.91 \pm 2.65 \times 10^5$ & $1.60 \pm 0.43 \times 10^1$\\
37.4  -- 52.8   & 44.4   &  8  & 11.4 & $2.43 \pm 0.86 \times 10^5$ & $9.25 \pm 3.27 \times 10^1$\\
52.8  -- 74.5   & 62.7   & 12  & 14.1 & $2.12 \pm 0.61 \times 10^5$ & $1.98 \pm 0.57 \times 10^2$\\
74.5  -- 105.3  & 88.6   & 5   &  5.2 & $5.53 \pm 2.47 \times 10^4$ & $1.38 \pm 0.62 \times 10^2$\\
105.3 -- 148.7  & 125.1  & 15  & 15.1 & $1.14 \pm 0.30 \times 10^5$ & $6.32 \pm 1.63 \times 10^2$\\
148.7 -- 210.0  & 176.7  & 7   & 7.0  & $3.76 \pm 1.42 \times 10^4$ & $4.73 \pm 1.79 \times 10^2$\\
210.0 -- 296.7  & 249.6  & 3   & 3.0  & $1.14 \pm 0.66 \times 10^4$ & $3.76 \pm 2.17 \times 10^2$\\
296.7 -- 419.1  & 352.6  & 0   &   0  & $0.00                     $ & $0.00$                     \\ 
419.1 -- 592.0  & 498.1  & 4   & 4.0  & $7.60 \pm 3.80 \times 10^3$ & $9.11 \pm 4.56 \times 10^2$\\
592.0 -- 836.2  & 703.6  & 1   & 1.0  & $1.35 \pm 1.35 \times 10^3$ & $3.46 \pm 3.46 \times 10^2$\\
836.2 -- 4702.3 & 1982.9 & 2   & 2.0  & $2.69 \pm 1.90 \times 10^3$ & $9.36 \pm 6.62 \times 10^3$\\
\hline
\end{tabular}
\end{table*}

\section{Comparisons with Other Radio Surveys: Spectra Indices}

As described in Section~2, we have used the NVSS survey (Condon et
al. 1998) to determine and correct any positional offset for the
brightest objects in this GMRT survey. Both the NVSS and the VLSS can
be used to determine spectral information for our sources. The main
purpose of this is to identify sources with unusual spectral indices,
particularly those with steep spectra indices, which are candidate
high redshift objects. We shall determine spectral indices between
150~MHz and 1.4~GHz ($\alpha_{150}^{1400}$) and also between 74~MHz
and 150~GHz ($\alpha_{74}^{150}$).

\subsection{The NVSS Survey (1.4~GHz)}

The National Radio Astronomy Observatory (NRAO) Very Large Array (VLA)
telescope sky survey (NVSS) covers the sky with $\delta>-40$ degrees,
at a frequency of 1.4~GHz, a spatial resolution of $45''$ and a
limiting source brightness of about 2.5~mJy~beam$^{-1}$ (Condon et
al. 1998). The NVSS has a broadly comparable sensitivity to the GMRT
observations presented here. For example, for $\alpha=0.8$, a source
just detected by the GMRT will also be just detected in the
NVSS. However, for steeper spectrum sources, with $\alpha>1$, this may
not be the case.

We can envisage three different permutations. Firstly, a source is 
detected with the GMRT and is also in the NVSS, second, a source is 
detected with the GMRT but is not in the NVSS, and third that the 
source is not detected with the GMRT but is in the NVSS. We will 
discuss sources in each of the three categories below.

There are a total of 227 NVSS sources within the GMRT survey area. The
majority are relatively faint, with only 45\% of the NVSS sources
having a flux $>5$~mJy, and 16\% with a flux $>20$~mJy.

We have already used the brightest 20 sources in NVSS and GMRT to
correct the positions of the GMRT sources. Due to the resolution of our 
observations, if the sources are within $30''$ of the GMRT positions then  
we consider them to be associated.

According to the possibilities listed above, we have the following:
\begin{enumerate}

\item There 81 matches between the GMRT and NVSS sources lists. The
 spectral index distribution of these sources is discussed in the
 following section. Unsurprisingly, all of the brightest GMRT sources
 are also detected in the NVSS.

\item There are 32 GMRT sources that are not detected in the NVSS. The
  brightest GMRT source that is not detected in the NVSS has a 150~MHz
  flux of 56.4~mJy. Some of these sources are possible USS
  sources. Assuming that a non-detection with the NVSS corresponds to
  a 1.4~GHz flux upper limit of 2.5~mJy, a GMRT flux of 56.4~mJy
  corresponds to $\alpha_{150}^{1400} > 1.42$, implying a steep spectrum
  source. We will discuss these sources more in the following section.

\item There are 114 NVSS sources with no GMRT counterparts. These are
  mostly faint sources, with an NVSS flux $<10$~mJy. The brightest
  NVSS source that is not detected by the GMRT has a flux of 50.3~mJy,
  implying a spectral index of $\alpha<-0.47$. Such a value is well
  within the bounds of spectral indices other surveys (see for example
  the comparison between observations at 610~MHz and 1.4~GHz presented
  by Garn et al. 2007).

\end{enumerate}

\subsection{The VLSS Survey (74~MHz)}

The VLA Low-Frequency Sky Survey (VLSS, see Cohen et al. 2007) is a
74~MHz continuum survey covering the entire sky north of $-30^\circ$
declination.  It uses the VLA in BnA and B-configurations and has
mapped the entire survey region at a resolution of $80''$ and with
an average {\sl rms} noise of 0.1~Jy~beam$^{-1}$ (Helmboldt et
al. 2008). The typical source detection limit for the VLSS is around
700~mJy. For a spectral index of $\alpha=0.8$, this VLSS flux limit
corresponds to a 150~MHz flux of 400~mJy. This indicates that only a
small fraction of the GMRT sources will be detected by the VLSS.

\begin{figure}
\includegraphics[scale=0.4]{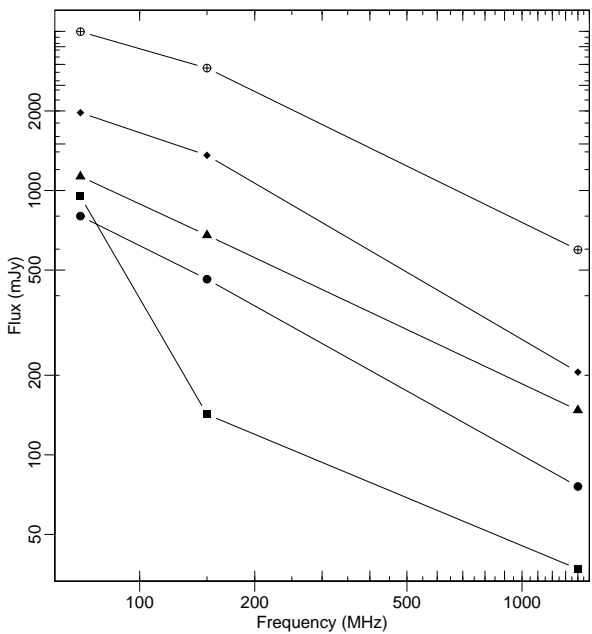}
\caption{The spectral indices for the 5 sources detected in the NVSS,
  VLSS and the GMRT, showing one object with an extreme spectral index
  between 74~MHz and 150~MHz (shown in the figure with a rectangular point)}
\label{nvssvlss}
\end{figure}

There are only 6 VLSS sources within the region of the GMRT survey. Of
these, 5 are detected by the GMRT (and also in the NVSS) and
are identified in Table~\ref{sources}, and they correspond to the
brightest GMRT sources. One of these sources (GMRTJ033152.9--100843),
detected with the GMRT and VLSS, seems to have a steep spectrum,
with $\alpha_{74}^{150}=2.5$ (Fig.~\ref{nvssvlss}).

In addition, one VLSS source (VLSS033315.89-085119.9) is detected in
the VLSS, but not by the GMRT (nor the NVSS). This source has a 74~MHz
flux density of 0.970~Jy. The non-detection with the GMRT suggests a
spectral index of $\alpha_{74}^{150}>5.2$, implying a very steep
spectrum source. It is worth noting though that there is a bright source 
close by and the background noise level is much higher around this object 
in the GMRT data.

We identify these two objects also as USS sources and candidate HzRGs, and
we will discuss these sources in the following section.

\subsection{Spectral Indices}
\label{spectral}

We are able to match 81 objects between our catalogue and
the NVSS. In addition, we can place constraints on those sources that
are detected with the GMRT but are not in the NVSS. Because we are
focusing on steep spectrum sources we will not consider those sources
in the NVSS that are not detected by the GMRT.

\begin{figure}
\includegraphics[scale=0.4]{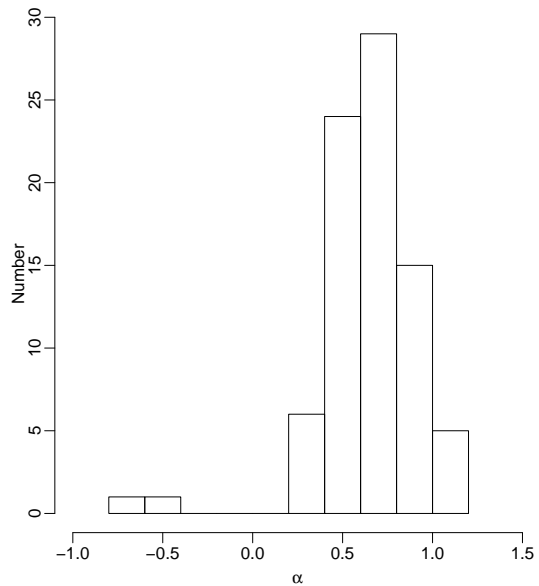}
\caption{Radio spectral index $\alpha$ between 150~MHz and 
1.4~GHz, for sources in the NVSS catalogue and our source catalogue.}
\label{nvss_spectral_index}
\end{figure}

The spectral index distribution of the matched sources from the NVSS
survey can be found in Fig.~\ref{nvss_spectral_index}. The
distribution has a median of $0.66$. 

\subsection{Candidate USS Sources}

We define USS sources as those having a spectral index $\alpha>
1.25$, either in the spectral range between 74~MHz and 150~MHz or that
between 150~MHz and 1.4~GHz.

Any GMRT source with a 150~MHz flux exceeding 38.5~mJy that is not detected in
the NVSS is deemed a candidate USS. A total of 3 sources fall in this
category. One source is detected in the VLSS but not with the GMRT and fulfils the criteria.
One of the sources, GMRTJ033152.9, is detected in all 3 surveys and shows a 
steep spectrum between 74~MHz and 150~MHz (see Table~1).

Consequently, we can identify a total of 5 candidate USS sources,
which are listed in Table~\ref{uss}. These objects will be the
subject of follow-up observations in a subsequent paper. As discussed in the introduction, 
a steep radio spectral index does not automatically imply a high redshift object 
and follow-up observations are necessary to confirm the nature of these sources.

\begin{table*}
\caption{Candidate USS sources. These are sources with a spectral 
index $\alpha>1.25$ in 
either the range 74--150~MHz or 150--1400~MHz.} 
\begin{tabular}{lccccl}\hline
Name & RA&  Dec & $\alpha_{74}^{150}$ & $\alpha_{150}^{1400}$ & Surveys / Detections\\
     & (J2000) & (J2000) &&&\\ \hline
GMRTJ033152.9--100843 & 03 31 52.97 & --10 08 43.7 & $2.5$ & $0.8$ & VLSS $\surd$, 
GMRT $\surd$, NVSS $\surd$\\
GMRTJ033318.7--085227 & 03 33 18.72 & --08 52 27.2 & -- & $>1.26$ & VLSS $\times$, GMRT $\surd$, NVSS $\times$\\
GMRTJ033216.0-083940 &  03 32 16.06 & --08 39 40.6 & -- & $>1.26$ & VLSS $\times$, GMRT $\surd$, NVSS $\times$\\
GMRTJ033326.8--084205 & 03 33 26.61 & --08 42 05.9 & -- & $>1.42$ & VLSS $\times$, GMRT $\surd$, NVSS $\times$\\
VLSSJ033315.8-085119 & 03 33 15.89 & --08 51 19.9 & $>5.2$&-- & VLSS $\surd$, GMRT $\times$, NVSS $\times$ \\
\hline
\end{tabular}
\label{uss}
\end{table*}

\section{Summary}

We have presented the results of a deep 150~MHz low frequency radio
map covering an area with a diameter of $>2^{\circ}$ in the region
around $\epsilon$ Eridani, and described the production of a catalogue
of 113 sources. The radio spectral index analysis of the sources in
the field made use of the NVSS (1.4~GHz) and VLSS (74~MHz).  Between
this 150~MHz dataset and the NVSS we have made reliable
identifications with the NVSS of 81 objects.

We have calculated the 150~MHz source flux distribution and find it to
be broadly comparable to that at other wavelengths.

We have made use of these observations, in conjunction with the NVSS
and VLSS to identify a number of USS sources. These observations have
shown that GMRT 150~MHz observations would can be effectively used to
search for USS sources. Of course LOFAR, when fully on-line, will be
considerably more effective at finding such sources than the GMRT.

These GMRT 150~MHz observations are clearly capable of detecting USS
sources, and we are in the process of following up some of the HzRG
candidates. It is also clear that a well designed 150~MHz survey,
in conjunction with observations at higher frequencies, could be 
powerful in detecting high redshift objects.

\section*{Acknowledgements}

SJG is supported by a STFC studentship. We thank the GMRT staff who
have made these observations possible. The GMRT is run by the National
Centre for Radio Astrophysics of the Tata Institute of Fundamental
Research. In particular, we would like to thank Ishwara Chandra C. H.
for his input during the observations and data reduction.  We thank
Timothy Garn, Dave Green, Derek Moss and Alastair Sanderson for helpful
discussions. We also wish to thank the referee, whose comments have improved 
this manuscript. The diagrams and statistics were generated using the R package.

\label{lastpage}
\end{document}